\documentclass[12pt,preprint]{aastex}

\begin{document}
\shorttitle{Optical Observations of SN 2007gi}

\shortauthors{Zhang, T. et al.}

\title{Optical Observations of the Rapidly Expanding \\Type Ia Supernova 2007gi}

\author{T.~Zhang\altaffilmark{1,2,3,4}, X.~Wang\altaffilmark{2,5,6},
W. Li\altaffilmark{5}, A. V. Filippenko\altaffilmark{5}, L.
Wang\altaffilmark{6}, X. Zhou\altaffilmark{1}, \\ P.~J.
Brown\altaffilmark{7}, J.~M.~Silverman\altaffilmark{5}, T.
N.~Steele\altaffilmark{5}, M.~Ganeshalingam\altaffilmark{5},
J.~Li\altaffilmark{2}, \\J.~Deng\altaffilmark{1},
T.~Li\altaffilmark{2}, Y.~Qiu\altaffilmark{1},
M.~Zhai\altaffilmark{1,4}, and R.~Shang\altaffilmark{2}}

\altaffiltext{1}{National Astronomical Observatories of China,
Chinese Academy of Sciences, Beijing 100012, China.}

\altaffiltext{2}{Physics Department and Tsinghua Center for
Astrophysics (THCA), Tsinghua University, Beijing, 100084, China.}

\altaffiltext{3}{LAM, CNRS, BP8, Traverse du Siphon, 13376 Marseille
Cedex 12, France.}

\altaffiltext{4}{Graduate University of Chinese Academy of Sciences.}

\altaffiltext{5}{Department of Astronomy, University of California,
Berkeley, CA 94720-3411.}

\altaffiltext{6}{Physics Department, Texas A\&M University, College
Station, TX 77843.}

\altaffiltext{7}{Department of Astronomy \& Astrophysics, 525 Davey
Laboratory, Pennsylvania State University, University Park, PA
16802.}

\begin{abstract}

We present optical photometry and spectra for the Type Ia supernova
(SN Ia) 2007gi in the nearby galaxy NGC 4036. SN 2007gi is
characterized by extremely high-velocity (HV) features of the
intermediate-mass elements (Si, Ca, and S), with expansion
velocities ($v_{\rm exp}$) approaching $\sim$15,500 km s$^{-1}$ near
maximum brightness (compared to $\sim$10,600 km s$^{-1}$ for SNe Ia
with normal $v_{\rm exp}$). SN 2007gi reached a $B$-band peak
magnitude of 13.25$\pm$0.04 mag with a decline rate of $\Delta
m_{15}(B)$(true) = 1.33$\pm$0.09 mag. The $B$-band light curve of SN
2007gi demonstrated an interesting two-stage evolution during the
nebular phase, with a decay rate of 1.16$\pm$0.05 mag (100
days)$^{-1}$ during $t = 60$--90 days and 1.61$\pm0.04$ mag (100
days)$^{-1}$ thereafter. Such a behavior was also observed in the HV
SN Ia 2006X, and might be caused by the interaction between
supernova ejecta and circumstellar material (CSM) around HV SNe Ia.
Based on a sample of a dozen well-observed $R$-band (or unfiltered)
light curves of SNe Ia, we confirm that the HV events may have a
faster rise time to maximum than the ones with normal $v_{\rm exp}$.

\end{abstract}


\section{Introduction}

Type Ia supernovae (SNe Ia) have been successfully utilized over the
past decade to measure the cosmic expansion history \citep{riess98,
per99} and explore the nature of dark energy
\citep[e.g.,][]{riess04, riess07, ast06, wv07}. The foundation for
the utility of SNe Ia as a cosmological tool is that some
distance-independent observables, such as the light-curve shape
parameters \citep{phi93, riess95, jha07, per97}, the color
parameters \citep{wlf03, wxf05}, or both \citep{tripp98, guy05},
have been found to correlate with their peak luminosity. These
empirical correlations can be used to calibrate the luminosities of
SNe Ia and measure their distances with a precision of $\sim$9\%.

A recent result suggests that the luminosity standardization of SNe
Ia can be improved to a level of $\sim$6\% by separating the SNe
into two groups based on a spectroscopic criterion [for details, see
\citet{wxf09a}, hereafter W09]. The expansion velocity ($v_{\rm exp}$)
of the SN ejecta is inferred from the blueshift of the absorption
minimum of Si`II $\lambda$6355, and the SNe are divided into one
group with normal $v_{\rm exp}$ (hereafter ``Normal" SNe Ia) and the
other group with high $v_{\rm exp}$  (hereafter ``HV"
SNe Ia)\footnote{See \citet{ben05} and \citet{bran06} for a
similar classification based on the velocity gradient and the
strength of the absorption features, respectively.}. W09 found that
the two groups have either different extinction laws or color
evolution. The cause for such a dichotomy might be related to the
properties of their progenitors. The HV SNe Ia are characterized by
stronger absorption features of intermediate-mass elements (IMEs, such
as Si, S, and Ca) at higher velocities as well as a red $B - V$ color
around maximum brightness.

SN 2002bo and SN 2006X are two of the
best-studied examples of this class \citep{ben04, wxf08a}. In
particular, the interstellar Na I~D lines were found to show
significant variations in the spectra of SN 2006X, likely pointing
to the presence of CSM produced by the progenitor system
\citep{pat07, chu08}. The possible detection of CSM around SN 2006X
is also supported by a flat evolution of the late-time light curve
\citep{wxf08a} and a detection of light echoes \citep{wxf08b, cy08}.
Similar variability of the Na I~D lines
was also observed in SNe 1999cl and 2007le \citep{blon09, sim09},
two other members of the HV SN Ia class. Contrasting with this,
multi-epoch, high-resolution spectral observations of SN 2007af, a
Normal SN Ia, do not reveal any significant signature of CSM
absorption \citep{sim07}. This raises the possibility that CSM might
be preferentially present for the SNe Ia in the HV class.

In this paper, we present optical observations of another member
of the HV SN Ia class, SN 2007gi. Our goal is to increase the sample
and understand the properties of well-observed HV SNe Ia.
Observation and data reductions are described in \S 2, while
\S 3 presents the $BVRI$ light curves, color curves,
reddening estimate, and an analysis of the rise time. Section~4
presents the spectral evolution. Our discussions and conclusions
are given in \S 5.

\section{Observations and Data Reductions}

SN 2007gi was discovered by K. Itagaki (IAUC 8864) on July 23.49
(UT dates are used throughout this paper)
at $\alpha$ = 12$^{\rm{h}}$01$^{\rm{m}}$23$^{\rm{s} }$.4, $\delta$ =
+61$^{\circ}$53$^{\prime}$33$^{\prime\prime}$.8 (J2000.0) in
the lenticular galaxy NGC 4036. An optical spectrum
taken on 2007 Aug. 4.90 revealed that SN 2007gi was a very young SN
Ia at a phase $\sim$10 days before maximum brightness \citep{hbc07},
with an expansion velocity measured from the absorption minimum of
Si~II $\lambda$6355 as $\sim$20,500 km s$^{-1}$. This velocity
is much higher than what is observed for a Normal SN Ia at a similar
phase,  but is typical for members of the HV SN Ia class, such as
SNe 1983G, 1997bp, and 2006X \citep{ben05, wxf08a}.

\subsection{Photometry}

Photometry of SN 2007gi was obtained primarily with the 0.8-m
Tsinghua-NAOC Telescope (TNT) located at NAOC
Xinglong Observatory\footnote{This telescope is operated by
Tsinghua University and the National Astronomical Observatories of
China (NAOC).}. This telescope is equipped with a $1340 \times 1300$
pixel back-illuminated CCD, with a field of view (FOV) of
$11.5^{\prime} \times 11.2^{\prime}$ (scale $\sim0.52^{\prime\prime}$
pixel$^{-1}$). Observations of SN 2007gi began on 2007 Aug. 8,
$\sim$6 days before $B$-band maximum, and continued for
the next 6 months.  SN 2007gi was also observed during its early
rising phase (from $t \approx -11$ to $-$7 days) with the
Ultraviolet/Optical Telescope [UVOT; \citet{roming05}] onboard the
{\it Swift} Observatory.

As shown in Figure 1, SN 2007gi is located just outside the central
bulge of NGC 4036. To remove the light contamination from the host
galaxy, we applied an image-subtraction technique before performing the
photometry. Template images of NGC 4036 were obtained on 2008 May
24, roughly 300 days after maximum brightness. To perform the image
subtraction, the image containing SN 2007gi is first geometrically
registered to the corresponding template image; the fluxes of
the foreground stars in the images are then scaled to the same level.
The point-spread functions (PSFs) of
these two images are convolved to match, and the template is then
subtracted from the SN images. We finally performed standard
PSF-fitting photometry to obtain the instrumental magnitudes for the
SN and the local standard stars with the IRAF\footnote{IRAF, the
Image Reduction and Analysis Facility, is distributed by the
National Optical Astronomy Observatories, which are operated by the
Association of Universities for Research in Astronomy (AURA), Inc.,
under cooperative agreement with the National Science Foundation.}
DAOPHOT package \citep{stet87}.

Transformations from the instrumental magnitudes to the standard
Johnson $UBV$ \citep{jimw66} and Kron-Cousins $RI$ \citep{cou81}
systems were established by observing on photometric nights a
series of \citet{lan92}
standard stars covering a wide range of air masses and colors.
A total of 4 photometric nights were used to
calibrate 8 local standard stars in the field of SN 2007gi. Table 1
lists their final calibrated $BVRI$ magnitudes \citep{lijz08}.
These local stars are then used to transform the instrumental
magnitudes of SN 2007gi to the standard $BVRI$ system, and the final
results of the photometry are listed in Table 2. The error bars (in
parentheses) include both the uncertainty in the calibration of the
local standard stars and the uncertainty of the instrumental
magnitudes due to photon noise and image subtraction.

The $B$- and $V$-band magnitude obtained by UVOT on three earlier
epochs are also listed in Table 2. The Swift/UVOT photometry was
reduced in a manner similar to that described in \citet{brown09}
and using the photometric zeropoints of \citet{pool08}.  In the
absence of template images of the host galaxy, regions around the
galaxy at the same distance from the nucleus as the SN were used to
estimate the underlying count rate and its uncertainty.

Given the sparse spectroscopic observations of SN 2007gi (\S 4), no
$S$- or $K$-corrections were applied to the photometry presented in
this paper.

\subsection{Spectroscopy}

Optical spectra of SN 2007gi were obtained primarily with the 3-m
Shane telescope at Lick Observatory using the Kast double
spectrograph \citep{ms93}. A spectrum was also obtained on 2007 Nov.
12 with the Low Resolution Imaging Spectrometer (LRIS; \citet{oke95}]
mounted on the 10~m Keck I telescope on Mauna Kea, Hawaii. The
journal of spectroscopic observations is given in
Table 3.

All of the spectra were reduced by standard IRAF routines. In
order to avoid contamination from the host galaxy, each spectrum
of the SN was extracted carefully, and the flux calibration was done
with spectra of standard stars observed on the same night at
similar air masses. The spectra were corrected for continuum
atmospheric extinction using mean extinction curves for Lick
Observatory, and telluric lines were removed from the data. During
the observations, the slit was always aligned near the parallactic
angle to avoid chromatic losses due to atmospheric dispersion
\citep{fil82}.

\section{Light Curves of SN 2007gi}

\subsection{Optical Light Curves}

The $BVRI$ light curves of SN 2007gi are presented in Figure 2. A
polynomial fit to the $B$-band light curve around maximum
brightness yields $B_{\rm max}$ = 13.25$\pm$0.04 mag on
JD 2,454,327.02$\pm$0.71 (2007 August 14.5).
This indicates that our observations
started from about $-$11 days and extended to +188 days with respect
to the $B$-band maximum.  The $V$-band light curve reached a peak
magnitude of 13.08$\pm$0.03 mag on JD 2,454,328.48$\pm$0.72, about
1.5 days after $t_{\rm max}$(B).

Using a template-fitting method [similar to that used in Hamuy et
al. (1996) but expanded to include more templates] to fit the $B$-
and $V$-band light curves, we derived $\Delta m_{15}(B)$ =
1.31$\pm$0.08 mag and $B_{\rm max} - V_{\rm max}$ = 0.17$\pm$0.05
mag. The $B - V$ color around the maximum is found to be slightly
redder than the unreddened loci \citep{wxf09b}, suggesting some
reddening toward SN 2007gi. The SALT2 method \citep{guy07} yields a
best-fit stretch factor of $s$ = 0.88$\pm$0.02, which corresponds to
$\Delta m_{15}$(B) = 1.40$\pm$0.10 from the transforming relation
between $s$ and $\Delta m_{15}$(B) by \citet{alt04}. The best-fit
templates from MLCS2k2 \citep{jha07} are also overplotted in Figure
2.  One can see that the light curves of SN 2007gi can be well fit
by both SALT2 and MLCS2k2.

In Figure 3, we compare the light curves of SN 2007gi with those
having similar $\Delta m_{15}(B)$ values:  SNe 1994D \citep{rich95,
pat96}, 1996X \citep{sal01}, 2002bo \citep{ben04}, 2002dj
\citep{pig08}, 2002er \citep{pig04, kot05}, 2004eo \citep{pas07},
and 2006X \citep{wxf08a}. Among these objects, SNe 2002bo, 2002dj, 2006X,
and 2007gi belong to the HV group, while SNe 1994D, 1996X, and 2004eo
belong to the Normal group. SN 2002er may be a transitional object linking
the Normal and HV groups (W09).

Compared to the Normal SNe Ia in the $B$ band, SN 2007gi appears
relatively brighter and declines slowly when entering the nebular
phase. For convenience of comparison, we introduce a quantity
$f(t_{\rm max})/f(t_{60})$, which is the flux ratio measured at
$t = 0$ and $t = 60$ days with respect to the $B$ maximum. The ratio
$f(t_{\rm max})/f(t_{60})$ is found to be 19.9$\pm$0.7 for SN 2007gi,
which is similar to that of SN 2006X (17.2$\pm$0.6) but noticeably smaller
than that of SN 1994D (27.3$\pm$0.6), SN 1996X (25.4$\pm$2.0), and SN 2004eo
(24.2$\pm$1.2). Thus, the HV objects seem to have systematically
smaller peak-to-tail contrast than the Normal objects. Close
inspection of Figure 3a also reveals that there may be a break in
the light-curve evolution for SN 2007gi at $t \approx 90$ days.
The decay rate $\beta$ is found to be 1.16$\pm$0.05 mag (100
days)$^{-1}$ during the period from $t \approx 60$ days to $t
\approx 90$ days, similar to that measured for SN 1984A and SN
2002bo at the same phase. The decay becomes steeper after $t
\approx 90$ days, with $\beta$ = 1.61$\pm$0.04 mag (100
days)$^{-1}$. Such a two-stage evolution also exists in SN 2006X,
but it is not unambiguously present in the cases of SNe 1984A and
2002bo due to poor photometric coverage in the nebular phase. Future
observations will be needed to determine whether this trend is
universal for all HV SNe Ia.

In contrary to the $B$-band light curve, the $VRI$-band light curves
of SN 2007gi do not exhibit a two-stage evolution in the nebular
phase (see Figure 3b--d). Neither the flux contrast between the tail
and peak nor the decay rate of the tail show significant differences
among the objects in the comparison. A noticeable difference is that
the secondary shoulder/maximum features at around +25 days in the
$VRI$ bands appear more pronounced in the HV objects than those in
the Normal ones (see Figures 3c and 3d). Part of this effect could be
caused by significant extinction which shifts the effective passband
to redder wavelengths [e.g., \citet{wxf08a}]. As the extinction
suffered by SN 2007gi is relatively low (\S 3.2), its prominent
secondary shoulder is likely to be mostly intrinsic. According to
\citet{kas06}, the variation of the secondary maximum of the
near-infrared (NIR) light curves is caused by the abundance
stratification in SNe Ia and/or the progenitor metallicity.

One feature in Figures 3(a)--(d), though not easily discernible
due to the crowdness of the data points, is that the HV SNe
2007gi, 2006X, and 2002bo seem to rise to the maximum at a faster rate
than the Normal objects. Such a trend was also
mentioned by \citet{pig08} in the qualitative comparison of SN
2002dj. A quantitative analysis of the rise time is given in \S 3.3.

\subsection{Color Curves and the Reddening}

Figure 4 presents the optical color curves of SN 2007gi ($B-V$, $V-R$,
$V-I$). Overplotted are the color curves of the other SNe in the
light-curve comparison in Figure 3.
All of the color curves are shifted to
match the observed values of SN 2007gi at $B$ maximum. For the
$B - V$ color, SN 2007gi
exhibited a similar color evolution as the other SNe in
the early phase ($t < 10$ days). At $t \approx 20$ days, it
has the reddest color of all the SNe.
After the red peak at $t \approx$ 25 days, SN 2007gi seems
to become progressively bluer than the other SNe in the
comparison, so by $t \approx 60$ days it has the bluest color.
Unfortunately, there is a gap in the range $t = 20$--60 days when SN 2007gi
was behind the Sun, so the detailed evolution during this phase is not clear.
The slope measured during $t = 60$--90 days is $-0.0151 \pm 0.0005$ mag
day$^{-1}$, steeper than the Lira-Phillips
relation [e.g., $-0.0108$ mag day$^{-1}$; \citet{lira95, phi99}].
This was also noticed by \citet{wxf08a} in the comparison study of SN
2006X. In fact, a steeper $B - V$ color-change slope than the
Lira-Phillips relation in the nebular phase is found
to be the case for most of the HV events (Wang, X., et al. 2009c, in preparation),
suggesting that the reddening derived for them from their tail colors
might be potentially biased toward a lower value. Despite the
relatively large differences seen in the $B - V$ color evolution,
the two classes of SNe Ia (HV and Normal) share
similar $V - R$ and $V - I$ color evolution (Figures 4b--c).

The Galactic extinction toward NGC 4036/SN 2007gi is $A^{\rm gal}_{V} =
0.078$ mag \citep{sch98}, corresponding to a color excess
of $E(B - V) = 0.024$ mag [adopting the standard reddening law of
\citet{car89}]. The peak $B - V$ color, corrected for the
Galactic reddening, is 0.15$\pm$0.05 mag. This corresponds to a
host-galaxy reddening of $E(B - V) = 0.20$ mag under the assumption that
the HV SNe Ia have intrinsic $B - V$ colors similar to those of the
Normal objects near maximum brightness (W09). Assuming $R_{V} = 1.56$,
as suggested by W09 for the HV SNe Ia, the extinction caused by the
dust within the host galaxy of SN 2007gi is $A_V \approx 0.31$ mag.

\subsection{The Late-Time HST Photometry}

The late-time evolution of SNe can provide valuable constraints on
the SN environments as well as their explosion models [e.g.,
\citet{mtl01, lwd02, sol04}]. The field of SN 2007gi was imaged with
the Wide Field Planetary Camera 2 (WFPC2) onboard the {\it Hubble Space
Telescope (HST)} on 2009 March 26, $\sim$600 days after discovery,
as part of {\it HST} program GO-10877 (PI: Weidong Li). A
cosmic-ray-split pair of observations was observed with each of the
F435W, F555W, F675W, and F814W filters. To locate SN 2007gi on these
WFPC2 images, we follow the technique detailed by \citet{lwd07}.
We use an $R$-band image of SN 2007gi taken with the Lick
Observatory 1-m Nickel telescope on 2007 Dec. 15, and identify four
stars that are present in the WFPC2 images. An astrometric
solution using these four stars yields a relatively poor precision of
$0\arcsec.43$ due to the faintness of the stars and the resolution
of the Nickel image. Nevertheless, within the 1$\sigma$ error
radius of the astrometric solution, we identify a single star which
we presume to be SN 2007gi. Figure 5 demonstrates this process.

To measure the magnitudes of SN 2007gi in the WFPC2 images, we
have used the ``HSTphot" package developed by \citet{dol00a}. After
performing the preprocessing suggested by the HSTphot manual,
including removal of cosmic rays, defects, bad pixels, and hot pixels,
we measure the magnitudes in the WFPC2 system. These magnitudes are
then converted to the ground-based broad-band $BVRI$ magnitudes using
the procedure described by \citet{dol00b}. We measure the following
magnitudes for SN 2007gi: $B$ = 24.49$\pm$0.10, $V$ =
24.20$\pm$0.13, $R$ = 24.63$\pm$0.28, and $I$ = 23.50$\pm$0.12.

SN 2007gi was found to decline by $\sim$11.1 mag in the $V$ band
by 590 days after the $B$ maximum. This flux drop is consistent with
the corresponding values observed in SN 1996X and SN 2003hv at
comparable phases \citep{sal01, lel09}. SN 1991T and 1998bu, which
showed evidence for the presence of light echoes \citep{sch94, cap01},
declined by $<10$ mag within 590 days since the maximum. In addition,
the $B - V$ color of SN 2007gi in the nebular phase is found to be
inconsistent with that predicted by the light-echo scenario. Thus. we
conclude that an interstellar echo may not be present in SN 2007gi.

\subsection{Rise Time}

The time spent by a SN from explosion to maximum
brightness is defined as the rise time $t_{\rm r}$, an
important parameter that links the explosion
physics to the progenitors of SNe~Ia \citep{leipin92}. The rise
time of a SN Ia is determined primarily by the rate at which the
interior energy is released and subsequently diffuses to the surface
of the supernova. Assuming that the luminosity of a SN Ia evolves
as an expanding fireball at very early phases \citep{gold98}, the
rise time $t_{\rm r}$ can be derived from the relation \citep{riess99}
\begin{equation}
L = \alpha (t + t_{\rm r})^{2},
\end{equation}
where $\alpha$ is the "speed" of the rise of the luminosity and $t$
is the time elapsed since maximum brightness.

In this subsection, we attempt to examine the difference in $t_{\rm
r}$ between HV and Normal SNe Ia. Instead of adopting the $B$-band
light curve \citep{riess99, stro07}, our analysis uses the $R$-band
light curve, as our earliest photometric data points are usually
obtained with a clear filter (used in our SN search) which is very
close to broad-band $R$ \citep{lwd03}. Only those data obtained
earlier than one week before the $R$-band maximum are used in the
fitting.

The computed rise times for SN 2007gi and several other
well-observed SNe Ia are plotted against the decline rate $\Delta
m_{15}(B)$ in Figure 6.  The SNe are divided into the HV (filled
circles) and Normal (open squares) groups according to the
classification scheme defined by W09. In each group, $t_{\rm r}$
correlates with $\Delta m_{15}(B)$: slower decliners have longer
rise times. The HV SNe Ia appear to have a faster (i.e., shorter)
rise time than the Normal ones at the same $\Delta m_{15}(B)$ with
the exception of SN 1994D, which lacks published photometry during
the very early phases. A similar trend was also noted by
\citet{pig08}. Assuming comparable amounts of nickel are synthesized
during the explosion, the rise time of a SN Ia is affected by the
efficiency of the photon diffusion, which depends on the opacity and
the location of the photosphere. A shorter $t_{\rm r}$ could
indicate that compared with a Normal SN Ia, the photosphere of a HV
SN Ia is less effectively heated by the $\gamma$-ray photons inside
the ejecta, or becomes optically thin at a faster pace due to the
more rapid expansion (see the discussion in \S 4.3).

\section{Optical Spectra}

We have 5 optical spectra of SN 2007gi obtained with the
3.0-m Shane telescope at Lick Observatory and the Keck I 10-m telescope
at the W. M. Keck Observatory, spanning from $t = -7.5$ to $t =
+153.5$~days since $B$-band maximum brightness. As shown
in Figure 7, the spectral evolution around maximum brightness
generally follows that of a normal SN~Ia but is characterized by
very broad and highly blueshifted absorption features at 3600~\AA\
(Ca~II H\&K), 6000~\AA\ (Si~II $\lambda$6355), and 8000~\AA\ (Ca~II
NIR triplet).

\subsection{Temporal Evolution of the Spectra}

In Figure 8, we compare the spectra of SN 2007gi with those of
SNe~Ia having similar $\Delta m_{15}$ values at four selected epochs
($t \approx -8$, $-1$, +6, and +3 months since the $B$ maximum). All
spectra have been corrected for the reddening and the redshifts of
the host galaxies. The reddening of each comparison SN~Ia is taken
from Table 1 in W09, and the $R_{V}$ value is assumed to be 1.57 for
the HV SNe Ia and 2.36 for the Normal objects according to the
analysis W09.

At $t \approx -7$ days (Figure 8a), the spectrum of SN 2007gi is
characterized by lines of singly ionized IMEs (Si, S, Mg, and Ca).
The absorption features of Si~II $\lambda$6355 and Ca II H\&K in SN
2007gi are deep and broad, comparable to those in the HV SN Ia
2002bo but significantly stronger than those in the Normal SNe
1994D, 2002er, and 2004eo. Around maximum brightness (Figure 8b), we
measured a line-strength ratio of Si~II $\lambda$5972 to Si~II
$\lambda$6355, known as $\mathcal{R}$(Si~II) \citep{nug95}, to be
$0.07\pm0.03$ for SN 2007gi. This is a relatively small
$\mathcal{R}$(Si~II) value and usually indicates an overluminous
event \citep{nug95}. Similar low values of $\mathcal{R}$(Si~II) are
observed for the other two HV objects, SNe 2006X and 2002bo.

By one week after maximum (Figure 8c), the strong absorptions of Si~II
and Ca~II lines are still present in the HV objects (SNe 2007gi, 2002bo,
2002dj, and 2006X), but the W-shape S~II lines have almost vanished.
At $t \approx 3$ months (Figure 8d), the spectra are dominated by iron
lines whose strength is similar in all of the objects, and a strong
Ca~II NIR absorption trough. SN 2006X was found to show a prominent
near-UV excess starting from $t = 30$ days, perhaps due to a local
light echo or an interaction of the SN ejecta with circumstellar
material \citep{wxf08b}. Close inspection of the $t \approx
89$ day spectrum of SN 2007gi reveals a flux excess at the blue end
when compared to the Normal SNe~Ia, consistent with its bluer $B -
V$ color at this phase (\S 3.2).

\subsection{Expansion Velocity of the Ejecta}

The blueshift of the absorption minima of some spectral features may
approximately trace the location of the photosphere in the early
phases of a SN. In this subsection, we examine some strong lines
such as Si~II $\lambda$6355 and Ca~II H \& K, as well as a weak line
S~II $\lambda$5640. The derived evolution of the expansion velocity
($v_{\rm exp}$)\footnote{The relativistic
effect is taken into account when calculating the ejecta
velocities of the SNe.} of the SNe is shown in Figure 9.
All of the velocities have been corrected for the redshifts of the
host galaxies.

SN 2007gi is one of the objects with the highest expansion
velocities in the comparison, with $v_{\rm exp}$ = 15,500 $\pm$ 300 km
s$^{-1}$ at $t = -1$ day. By comparison, the typical value of $v_{\rm exp}$
for a Normal SN Ia is 10,600 $\pm$ 400 km s$^{-1}$ at this epoch (W09).
The velocities inferred from the Ca~II H $\&$ K and S~II $\lambda$5640
lines are about 21,500 km s$^{-1}$ and 12,800 km s$^{-1}$ near
maximum, respectively, which are also significantly higher than
those of the Normal SNe~Ia.

These high $v_{\rm exp}$ values of the IMEs (Si, S, and Ca) can often be
interpreted as being caused by a density/abundance enhancement in the
outer ejecta of the SNe \citep{len01, ben04}. Several scenarios have
been proposed to account for such an enhancement. The metallicity
effect was examined by \citet{ben04}, who found that increasing the
metallicity in the C+O layers by a factor of 10 is still far from
explaining the high $v_{\rm exp}$ seen in SN 2002bo. The delayed
detonation models were proposed to explain the HV features in SN
1984A and SN 2002bo \citep{len01, ben04}, as intense
nucleosynthesis in the outer layers results in an enhancement of the
density of the IMEs at higher velocities, which subsequently
increases the opacity in the outer regions and results in a
photosphere that is moving at a high velocity. An interaction between
the SN ejecta and CSM could produce the high-velocity Ca~II lines
\citep{ger04} and may also affect the Si~II lines \citep{tana06}.
Detailed analysis of the reasons for the HV features is beyond
the scope of this paper.

\subsection{The Photospheric Temperature}

As discussed by \citet{nug95}, \citet{bong06}, and \citet{hachinger06},
the flux ratio $\mathcal{R}$(Si~II) between
Si~II $\lambda$5972 and Si~II $\lambda$6355 is
an indicator of the photospheric temperature and the luminosity of a
SN~Ia: a value of $\mathcal{R}$(Si~II) means a more luminous SN~Ia with
a higher photospheric temperature. The unusually low $\mathcal{R}$(Si~II)
value of $\sim$0.07 for SN 2007gi at $t = -1$ day thus suggests a
rather high photospheric temperature. On the other hand, the strength of
Si~III $\lambda$4560 (actually Si~III $\lambda\lambda$4553, 4568) has
been proposed as another photospheric temperature tracer \citep{ben04},
and is stronger for a SN~Ia having a higher temperature. The
Si~III $\lambda$4560 line is weak in SN 2007gi, suggesting a rather
low photospheric temperature. The two temperature indicators thus
provide conflicting results for SN 2007gi.

In an attempt to further constrain the photospheric temperature in SN
2007gi, we used the parameterized resonance scattering
synthetic-spectrum code SYNOW \citep{fish99, bran05} to fit the
near-maximum spectra of SN 2007gi. The basic input parameters are
taken from Table 1 in \citet{bran05}, with the blackbody temperature
$T_{\rm bb}$ and the velocity at the photosphere $v_{\rm phot}$ as free
parameters. Figure 10 shows our best SYNOW fit for the $t = -1$ day
spectrum, with $v_{\rm phot}$ = 15,500 km s$^{-1}$ and $T_{\rm bb} =
13,500$~K. The high $v_{\rm phot}$ is consistent with the high
$v_{\rm exp}$ measured directly from the spectrum (\S 4.2). The blackbody
temperature $T_{\rm bb}$, on the other hand, is rather typical of other
SNe~Ia at this phase. For example, \citet{bran05} estimated
$T_{bb} = 13,000$~K for the Normal SN Ia 1994D at $t = +1$ day, while
our best SYNOW fit to a spectrum of the Normal SN Ia 2004eo at $t =
+1$ day yields $T_{\rm bb} = 10,500$~K.

Our SYNOW analysis thus suggests that $\mathcal{R}$(Si~II) is not
a good photospheric temperature indicator for the HV SNe~Ia.
The exact cause for this is not clear. One possibility is
contamination by Fe~II and Fe~III lines. \citet{bong08} suggested
that the Fe lines might weaken the Si~II $\lambda$5972 absorption line
even at low $T_{\rm bb}$.

\section{Discussion and Conclusions}

In this paper, we present optical photometry and spectroscopy
of SN 2007gi, a SN~Ia with an extremely high expansion
velocity measured from IMEs (Si, Ca, and S).
We also conduct a comparison study for a sample of SNe~Ia with high
and normal expansion velocities (HV and Normal, respectively) to
investigate differences in their photometric and spectroscopic
behaviors.

The $B$-band light curve of SN 2007gi shows a two-stage evolution: a
decay rate of $\beta = 1.16 \pm 0.05$ mag (100~days)$^{-1}$ during $t
\approx 60$--90 days and $\beta = 1.61 \pm 0.05$ (100~days)$^{-1}$
thereafter. The latter decay rate is similar to those observed in
Normal SNe Ia. This two-stage evolution is also present in the HV SN
Ia 2006X. More late-time observations of the HV objects are
necessary to establish whether this is a universal property. The $B
- V$ color of SN 2007gi is found to evolve at a faster slope than
that of Normal SNe Ia during the nebular phase, a trend that is also
observed for the HV SN Ia 2006X \citep{wxf08a} and most of the
other HV events (Wang, X., et al. 2009c, in preparation). SN 2007gi was detected at $t
\approx 600$ days after $B$ maximum in {\it HST}/WFPC2 images.  The
magnitudes and colors at this very late-time epoch suggest that SN
2007gi may not be contaminated by a light echo on interstellar
scales.

Using a dozen well-observed SNe Ia, we confirm the previous claim
that HV SNe Ia tend to have a faster rise time than Normal SNe Ia
\citep{pig08}: at the same value of $\Delta m_{15}$, the rise time of
the HV objects is shorter than that of the Normal objects by 1--2 days.

The spectral evolution of SN 2007gi is characterized by high
expansion velocities measured from the absorption lines. The value of
$v_{\rm exp}$ is measured to be 15,500 km s$^{-1}$ at $t = -1$ day,
$\sim$50\% higher than what is observed for Normal SNe Ia.
A small flux ratio $\mathcal{R}$(Si~II) = 0.07$\pm$0.03 is observed,
which traditionally means a hot photospheric temperature in the
SN ejecta. Our synthetic
spectral analysis using the SYNOW code, on the other hand,
suggests that SN 2007gi has a typical photospheric temperature
compared with other Normal SNe Ia.

The two-stage evolution in the $B$-band light curve of SN 2007gi may
suggest that an additional energy source besides radioactive decay
plays a role in the nebular phase.  Dust scattering (a light echo)
seems unlikely because the tail luminosity would have remained
nearly constant for a long time, which is not the case for SN
2007gi.  Nevertheless, the possibility of a local light echo on a
small scale (due to CSM dust) cannot be completely ruled out.
Another potential energy source is interaction with CSM; detection
of variable sodium lines in the spectra of some HV events provides
some evidence for this scenario. Assuming that the CSM lies at a
distance of $\sim10^{17}$ cm from the supernova, it is expected that
the outermost ejecta will begin to interact with the CSM at $\sim$30
days after the $B$ maximum. This naturally accounts for a flatter
evolution seen in the $B$-band light curve of the HV SNe Ia after $t
= 40$ days. The difficulty with this scenario is the lack of
convincing observational evidence for stripped material, such as
low-velocity H$\alpha$ emission in the nebular spectra [e.g.,
\citet{mat05}, but see \citet{leo08} for alternative explanations
regarding the observed lack of hydrogen], and the lack of spectral
evidence in support of ongoing CSM interaction.

A possible explanation for the HV features observed in the
HV SNe Ia is that there are density enhancements of the IMEs
in the outer layers of the ejecta of the HV objects
(due to a metallicity effect,
delayed detonations, or CSM interaction); the
$\gamma$-ray heating is less effective, and the photosphere is
formed at an outer layer with a higher expansion velocity compared
to the Normal objects.

\acknowledgments

We thank Peter Nugent for help with the SYNOW fitting. This study is
supported by the Chinese National Natural Science Foundation through
grants 10673007, 10673012, and 10603006, and by the China-973 Program
2009CB824800. A.V.F.'s group at U.C. Berkeley is grateful for NSF
grants AST-0607485 and AST-0908886, the TABASGO Foundation, and US
Department of Energy grants DE-FC02-06ER41453 (SciDAC) and
DE-FG02-08ER41563.  Financial support for this work was also provided
by NASA through grant GO-10877 from the Space Telescope Science
Institute, which is operated by Associated Universities for Research
in Astronomy, Inc., under NASA contract NAS 5-26555.  The work of
L.W. is supported by NSF grant AST-0708873. J.D. and Y.Q. are
supported by the NSFC grant No. 10673014 and the China-973 Program
2009CB824800.  Some of the data presented herein were obtained at the
W. M. Keck Observatory, which is operated as a scientific partnership
among the California Institute of Technology, the University of
California, and NASA; the observatory was made possible by the
generous financial support of the W. M. Keck Foundation.  We are
grateful to the staffs at the Lick and Keck Observatories for their
assistance with the observations.

\clearpage

\begin{figure}[ht]
\centering
\includegraphics[angle=0,width=150mm]{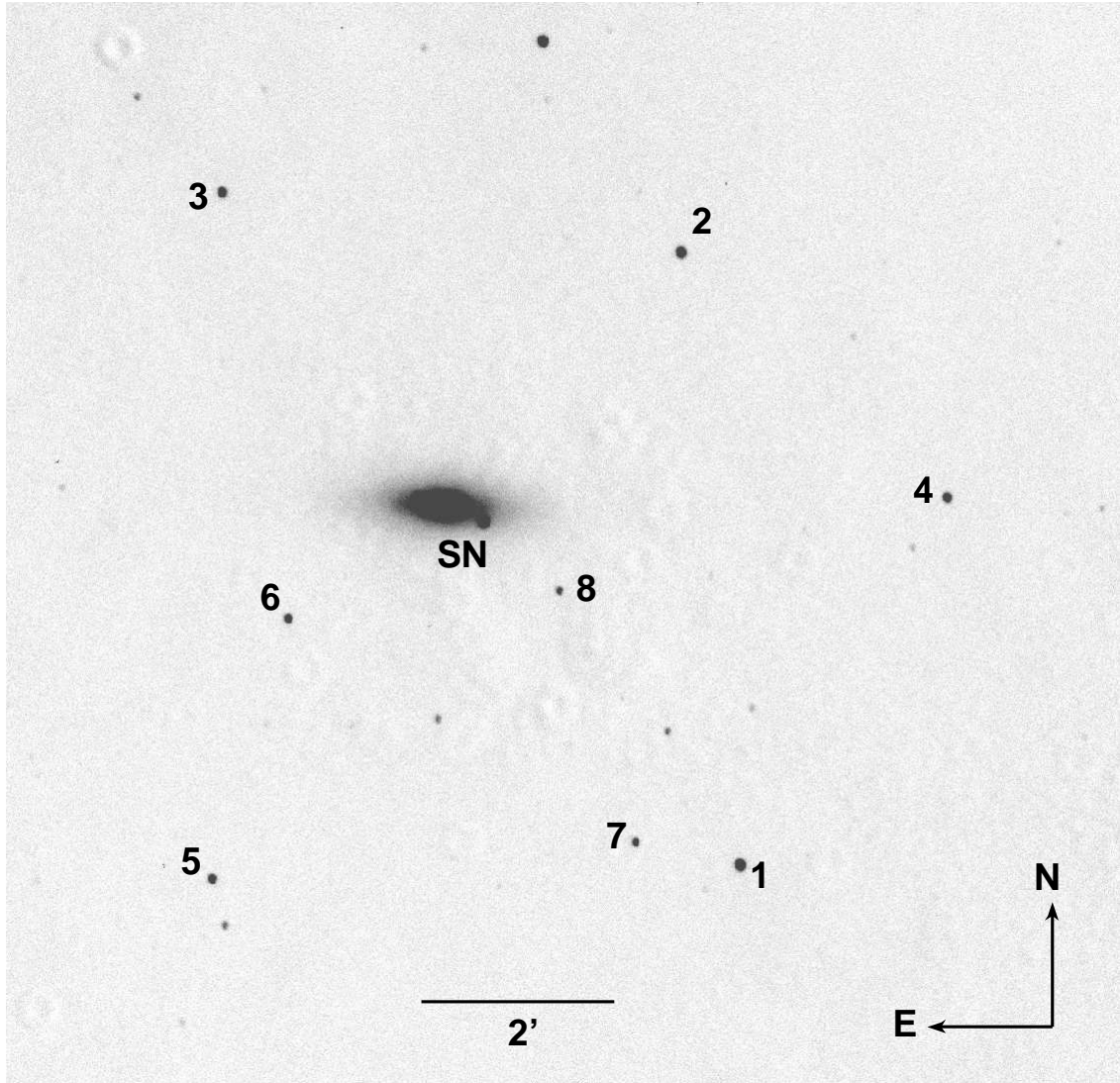}
\caption{SN 2007gi in NGC 4036, in an $R$-band image taken by
the 0.8-m TNT on 2007 August 14. The supernova and local reference
stars are marked. North is up and East is to the left.}
\end{figure}

\begin{figure}[ht]
\centering
\includegraphics[angle=0,width=150mm]{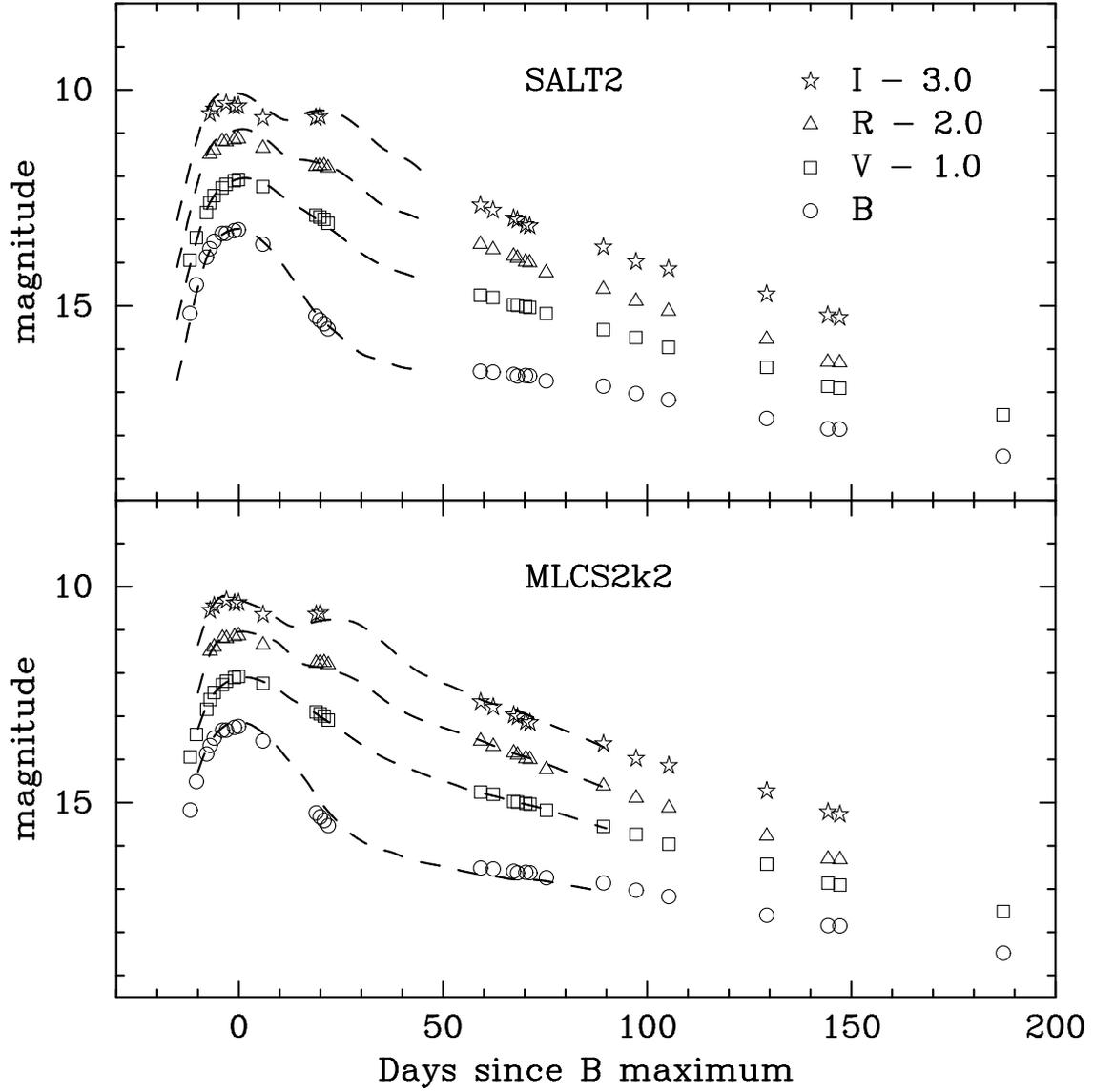}
\caption{$BVRI$ light curves of SN 2007gi, overplotted with the
best-fit SALT2 and MLCS2k2 templates.}
\end{figure}

\begin{figure}[ht]
\centering
\includegraphics[angle=0,width=150mm]{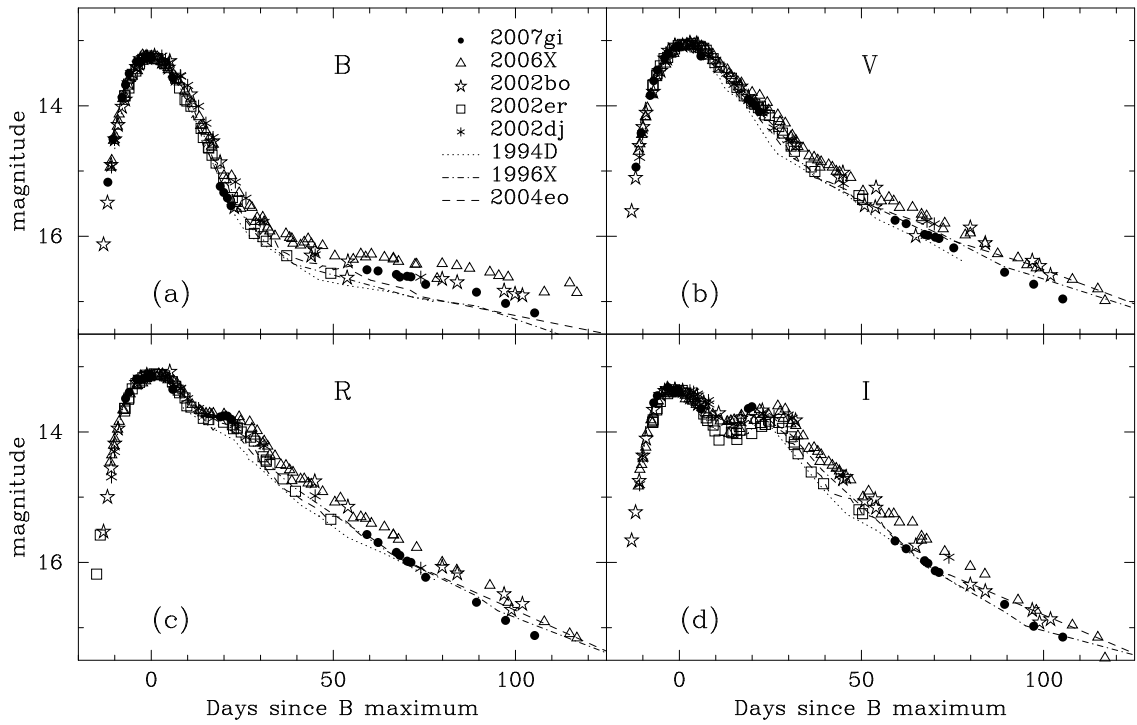}
\caption{$BVRI$ light curves of SN 2007gi, in comparison with those
of SNe 1994D, 1996X, 2002bo, 2002dj, 2002er, 2004eo, and 2006X. See
text for the references.}
\end{figure}

\begin{figure}[ht]
\includegraphics[angle=0,width=150mm]{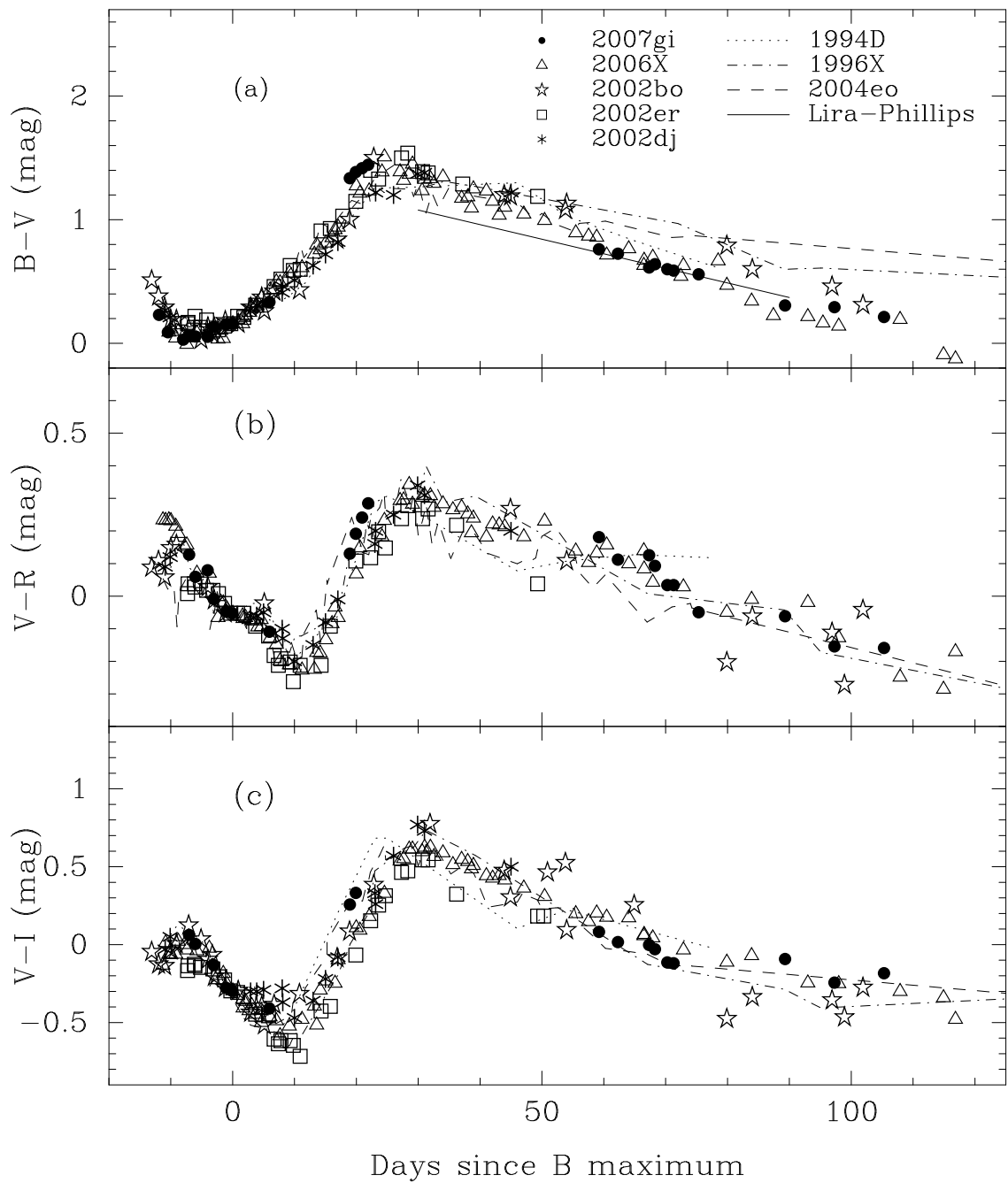}
\caption{$B-V$, $V-R$, and $V-I$ color curves of SN 2007gi compared
with those of SNe 1994D, 1996X, 2002bo, 2002er, 2002dj, 2004eo, and
2006X. All of the comparison SNe have been dereddened. The
solid line in the $B - V$ panel shows the unreddened
Lira-Phillips locus. The data sources are cited in the text.}
\end{figure}

\begin{figure}[ht]
\includegraphics[angle=0,width=150mm]{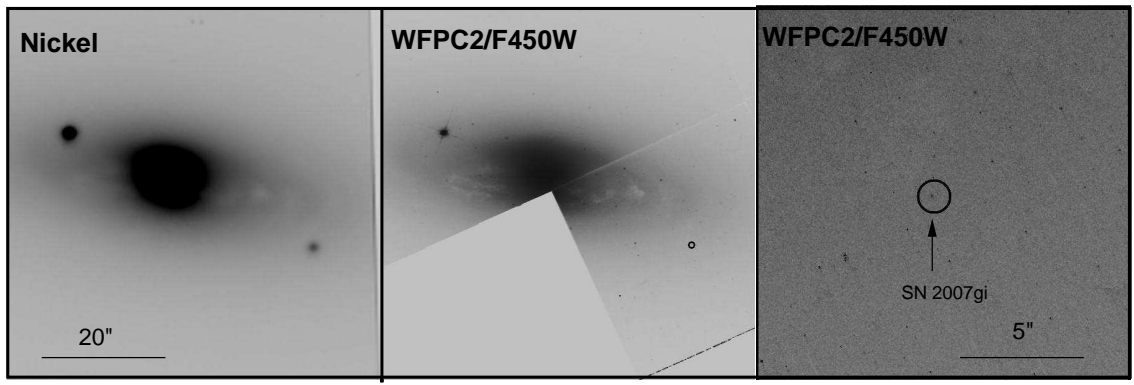}
\caption{Identification of SN 2007gi in the {\it HST}/WFPC2 images.
The left and middle panels show a $60\arcsec \times60\arcsec$
region of an $R$-band image taken with the Lick Observatory 1-m
Nickel telescope, and the {\it HST}/WFPC2 F450W image, respectively, after
the images have been astrometrically registered. The position of
SN 2007gi in the Nickel image is marked as a circle in the middle
panel. The right panel shows a $15\arcsec \times 15\arcsec$ region
around the site of SN 2007gi in the WFPC2/F450W image. The location
of SN 2007gi is marked with a circle having a radius that is three
times the astrometric precision. There is a single object
near the center of the circle which we identify as SN 2007gi.}
\end{figure}

\begin{figure}[ht]
\centering
\includegraphics[angle=0,width=150mm]{f6.ps}
\caption{The $R$-band rise time of SN 2007gi and other SNe Ia. The sample is
divided into two groups: HV (solid circles) [SNe 2002bo \citep{ben04},
2003W \citep{hick09}, 2006le \citep{hick09}, 2006X \citep{wxf08a},
 2007gi (this paper)] and Normal (open squares)
 [SNe 1990N \citep{lira98}, 1994D \citep{rich95, pat96},
 2001el \citep{kris03}, 2002er \citep{pig04}, 2003cg \citep{er06},
 2004eo \citep{pas07}, 2005cf \citep{wxf09b}, 2007af \citep{hick09} and SNe 1999cp, 2001en, 2001ep, 2002dj, 2002eb, 2003fa, 2006gr (Ganeshalingam et al. 2009, in preparation; Wang X. et al. 2009c, in preparation)].}
\end{figure}

\begin{figure}[ht]
\centering
\includegraphics[angle=0,width=150mm]{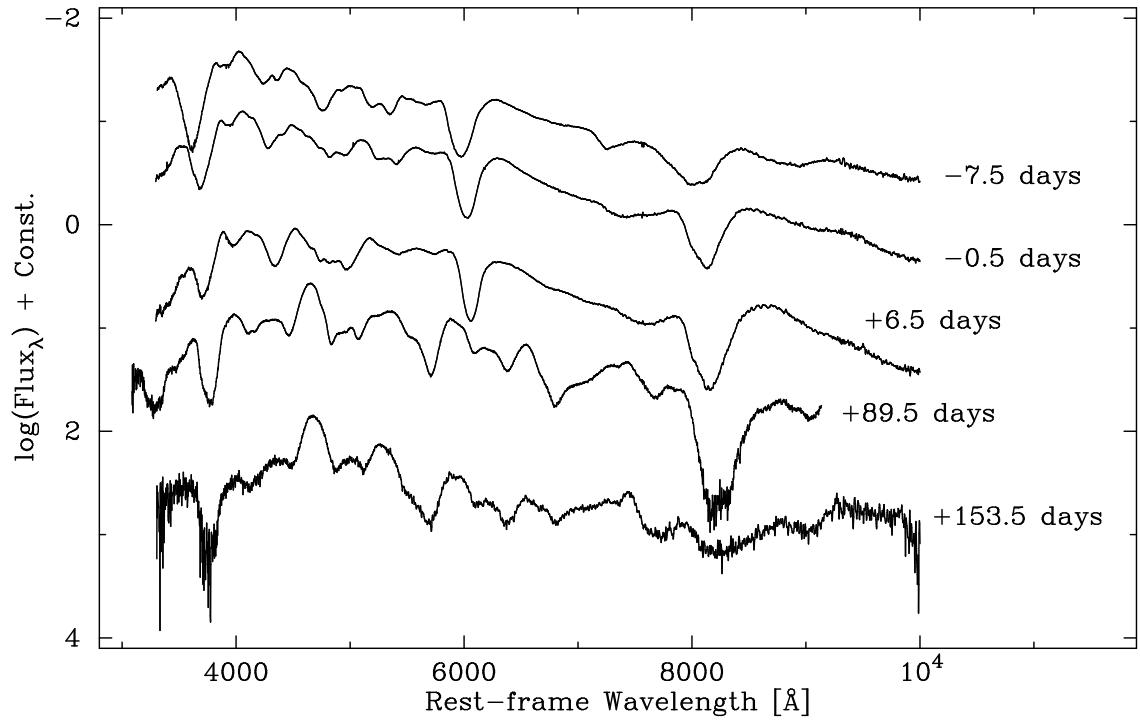}
\caption{Optical spectral evolution of SN 2007gi. All of the spectra
have been corrected for the heliocentric redshift of NGC 4036 ($v =
1445$ km s$^{-1}$) but not reddening. The spectra were arbitrarily
shifted in the vertical direction for clarity.}
\end{figure}

\begin{figure}[ht]
\centering
\includegraphics[angle=0,width=150mm]{f8.ps}
\caption{Spectra of SN 2007gi at four selected epochs ($t \approx
-8$, $-1$, +6, and +3 months from $B$-band maximum, from top to
bottom), compared to those of SNe 1994D \citep{pat96}, 1996X \citep{sal01},
2002bo \citep{ben04}, 2002dj \citep{pig08}, 2002er \citep{kot05}, 2004eo \citep{pas07},
and 2006X \citep{wxf08a, yama09} at the same phase. All of the spectra have
been corrected for reddening and host-galaxy redshift,  and shifted in
the vertical direction for clarity.}
\end{figure}

\begin{figure}[ht]
\centering
\includegraphics[angle=0,width=130mm]{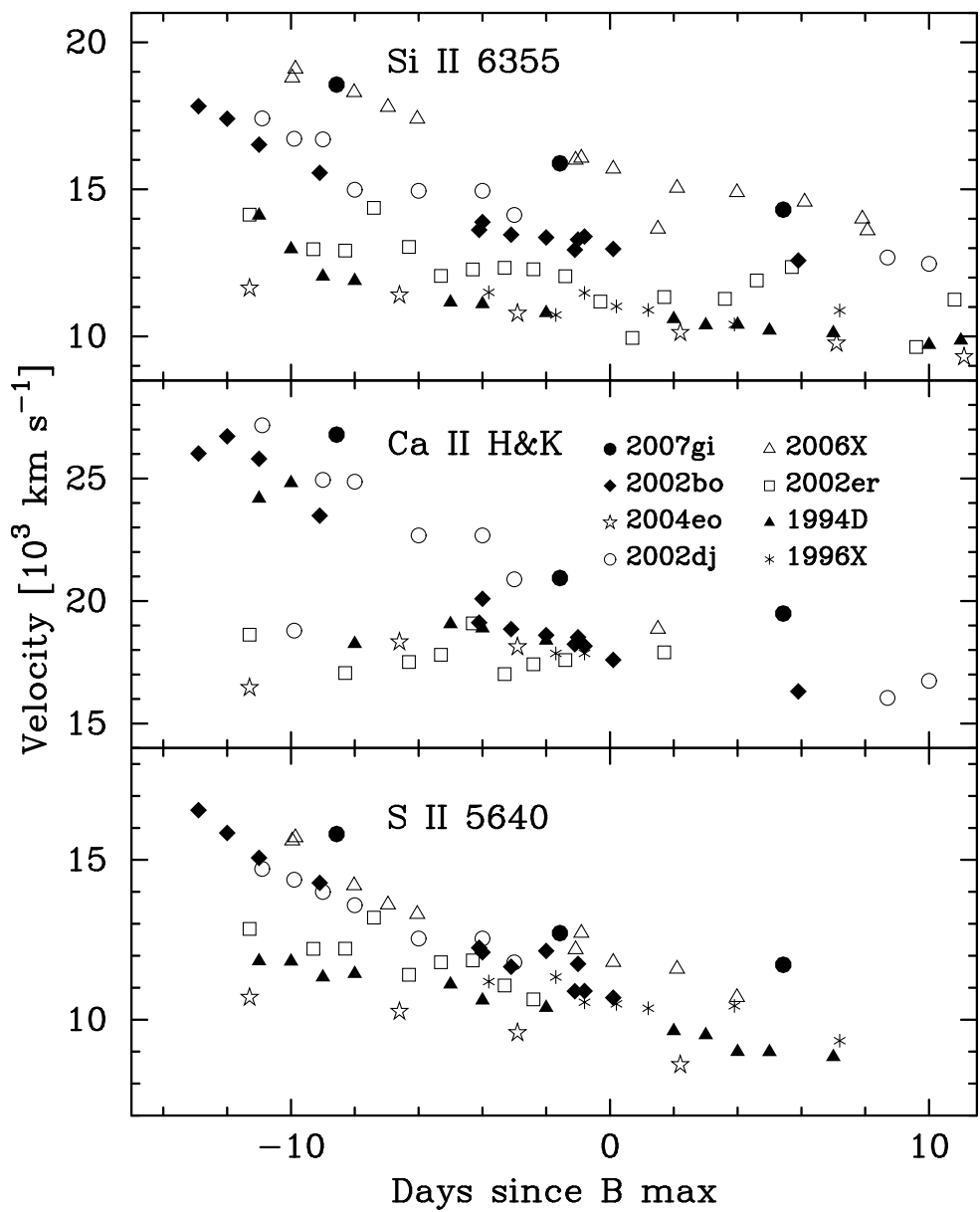}
\caption{ Evolution of the expansion velocity of SN 2007gi as
measured from the absorption minima of Si~II $\lambda$6355, Ca~II
H\&K, and S~II $\lambda$5640, compared with those of
SNe 1994D, 1996X, 2002bo, 2002dj, 2002er, 2004eo, and 2006X (see
the caption of Figure 8 for references).}
\end{figure}

\begin{figure}[ht]
\centering
\includegraphics[width=130mm]{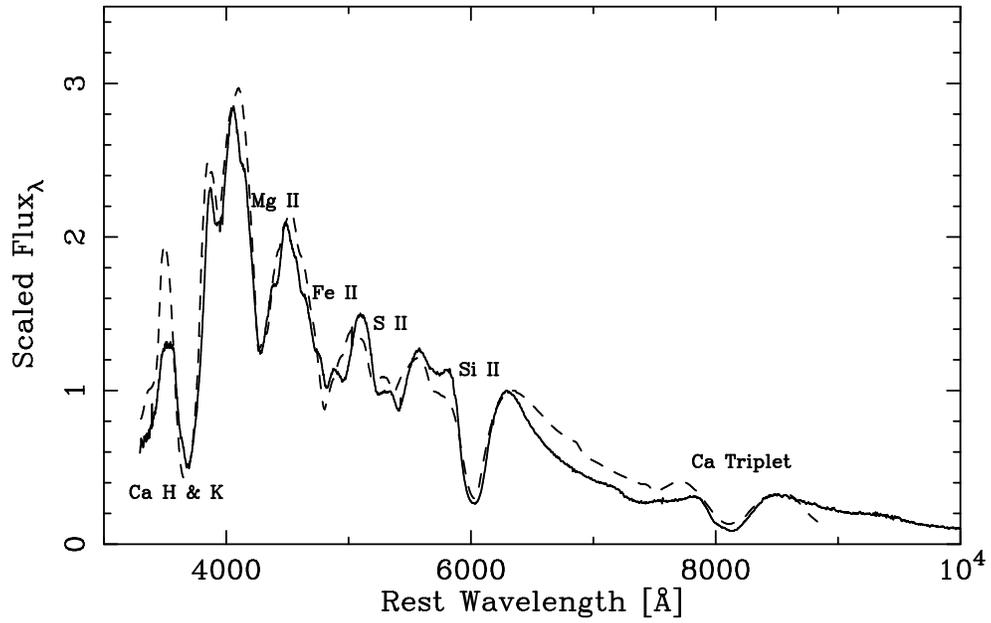}
\caption{The near-maximum spectrum of SN 2007gi (at $t = -0.6$ day),
overplotted with a synthetic spectrum (dashed lines) obtained by
the SYNOW code. See text for details. }
\end{figure}

\clearpage

\begin{table}
\caption{Magnitudes of Photometric Standards in the SN 2007gi Field}
{\small
  \begin{tabular}{lcccccc}
  \tableline\tableline
Star\tablenotemark{a}  & $V$ & $B-V$ & $U-B$ & $V-R$ & $V-I$\\
\tableline
1 & 13.860(002) & 0.823(003) & 2.392(016) & 0.475(002) & 0.726(002)\\
2 & 14.233(002) & 0.860(003) & 2.702(020) & 0.482(003) & 0.697(002)\\
3 & 14.833(002) & 0.742(004) & 2.399(022) & 0.395(004) & 0.585(004)\\
4 & 15.375(004) & 1.413(008) & 3.367(067) & 0.929(005) & 1.363(004)\\
5 & 15.298(003) & 0.800(006) & 2.532(032) & 0.443(005) & 0.648(005)\\
6 & 15.408(004) & 0.762(007) & 2.538(034) & 0.451(005) & 0.654(006)\\
7 & 15.869(005) & 0.878(009) & 2.655(045) & 0.522(007) & 0.783(007)\\
8 & 16.175(006) & 0.781(010) & 2.474(047) & 0.410(009) & 0.611(010)\\
  \tableline
  \end{tabular}
  \tablenotetext{a}{See Figure 1 for a chart of SN 2007gi and the comparison stars.}
}
\end{table}

\begin{table}
\caption{Optical Photometry of SN 2007gi}
{\footnotesize
  \begin{tabular}{lcrcccc}
  \tableline\tableline
  UT Date & JD $-$ 2,450,000 & Phase\tablenotemark{a} & $B$~(mag) &  $V$~(mag) & $R$~(mag) & $I$~(mag) \\  \tableline
2007 Aug 3 & 4316.17 & -10.85& 15.170(0.048) & 14.940(0.055) & \nodata       & \nodata     \\
2007 Aug 4 & 4317.70 & -9.32& 14.510(0.050) & 14.420(0.045) & \nodata       & \nodata     \\
2007 Aug 7 & 4320.18 & -6.84 & 13.870(0.052) & 13.840(0.053) & \nodata       & \nodata     \\
2007 Aug 8 & 4321.02 & -6.00 & 13.678(0.016) & 13.615(0.029) & 13.488(0.021) & 13.553(0.035)\\
2007 Aug 9 & 4322.03 & -4.99 & 13.505(0.015) & 13.450(0.028) & 13.391(0.034) & 13.446(0.029)\\
2007 Aug 11& 4324.03 & -2.99 & 13.323(0.021) & 13.269(0.048) & 13.190(0.018) & \nodata \\
2007 Aug 12 & 4325.02 & -2.00 & 13.321(0.023) & 13.189(0.028) & 13.197(0.017) & 13.320(0.014)\\
2007 Aug 14 & 4327.01 & -0.01 & 13.259(0.037) & 13.106(0.052) & 13.153(0.023) & 13.383(0.023)\\
2007 Aug 15 & 4328.01 & +0.99 & 13.237(0.028) & 13.077(0.029) & 13.130(0.033) & 13.374(0.032)\\
2007 Aug 21 & 4334.00 & +6.98 & 13.571(0.023) & 13.238(0.018) & 13.347(0.067) & 13.647(0.095)\\
2007 Sep 3 & 4347.02 & +20.00 & 15.236(0.024) & 13.899(0.015) & 13.769(0.028) & 13.642(0.018)\\
2007 Sep 4 & 4347.99 & +20.97 & 15.330(0.029) & 13.944(0.018) & 13.753(0.045) & 13.612(0.033)\\
2007 Sep 5 & 4348.99 & +21.97 & 15.414(0.028) & 13.997(0.025) & 13.756(0.036) &  \nodata\\
2007 Sep 6 & 4349.99 & +22.97 & 15.533(0.047) & 14.089(0.041) & 13.804(0.050) &  \nodata\\
2007 Oct 13 & 4387.30 & +60.28 & 16.514(0.020) & 15.753(0.012) & 15.572(0.017) & 15.671(0.026)\\
2007 Oct 16 & 4390.36 & +63.34 & 16.532(0.035) & 15.807(0.014) & 15.695(0.027) & 15.791(0.036)\\
2007 Oct 21 & 4395.39 & +68.37 & 16.586(0.020) & 15.972(0.013) & 15.846(0.019) & 15.973(0.035)\\
2007 Oct 22 & 4396.34 & +69.32 & 16.626(0.021) & 15.986(0.013) & 15.893(0.026) & 16.015(0.031)\\
2007 Oct 24 & 4398.34 & +71.32 & 16.613(0.031) & 16.014(0.012) & 15.980(0.038) & 16.128(0.047)\\
2007 Oct 25 & 4399.36 & +72.34 & 16.622(0.033) & 16.034(0.019) & 16.000(0.049) & 16.154(0.050)\\
2007 Oct 29 & 4403.40 & +76.38 & 16.736(0.059) & 16.179(0.081) & 16.228(0.081) & \nodata\\
2007 Nov 12 & 4417.38 & +90.36 & 16.857(0.016) & 16.551(0.020) & 16.612(0.017) & 16.643(0.056)\\
2007 Nov 20 & 4425.36 & +98.34 & 17.028(0.014) & 16.735(0.010) & 16.889(0.014) & 16.979(0.027)\\
2007 Nov 28 & 4433.36 & +106.34 & 17.174(0.033) & 16.961(0.022) & 17.120(0.020) & 17.145(0.035)\\
2007 Dec 22 & 4457.36 & +130.34 & 17.606(0.049) & 17.425(0.053) & 17.771(0.041) & 17.729(0.057)\\
2008 Jan 6 & 4472.36 & +145.34 & 17.845(0.023) & 17.861(0.032) & 18.301(0.044) & 18.218(0.086)\\
2008 Jan 9 & 4475.26 & +148.24 & 17.852(0.057) & 17.907(0.058) & 18.315(0.065) & 18.274(0.140)\\
2008 Feb 18 & 4515.29 & +188.27 & 18.480(0.030) & 18.520(0.126) & \nodata & \nodata\\
  \tableline
  \end{tabular}
  \tablenotetext{a}{Relative to the epoch of $B$-band maximum (JD = 2,454,327.02).}
}
\end{table}

\begin{table}
\caption{Parameters of SN 2007gi}
{\small
  \begin{tabular}{cccccc}
  \tableline\tableline
  Stretch & $\Delta m_{15}$ & $M_{V}^{max}$\tablenotemark{a} & Explosion velocity\tablenotemark{b} & $E(B-V)$ & Temperature\tablenotemark{c}\\
  & mag & mag & km s$^{-1}$ & mag & K\\ \tableline
0.88 $\pm$ 0.02 & 1.31 $\pm$ 0.09  & -19.26 $\pm$ 0.10 & 15, 500 $\pm$ 300 & 0.17 $\pm$ 0.04 & 11, 500\\
  \tableline
  \end{tabular}
  \tablenotetext{a}{Using 24.6 Mpc \citep{tully88} as the distance, $H_{0}$ = 71 km s$^{-1}$ Mpc$^{-1}$.}
  \tablenotetext{b}{Based on Si~II $\lambda$6355 absorption minimum at $B$ maximum.}
  \tablenotetext{c}{The blackbody temperature given by SYNOW at $B$ maximum.}
  }
\end{table}

\begin{table}
\caption{Optical Spectroscopy of SN 2007gi}
  \begin{tabular}{lcrlc}
  \tableline\tableline
  UT Date & JD $-$ 2,450,000 & Phase\tablenotemark{a} & Range(\AA) & Resolution(\AA)\tablenotemark{b}\\  \tableline
2007 Aug 7 & 4319.50  & -7.52 & 3300--10,000 &5-12\\
2007 Aug 14 & 4326.50 & -0.52 & 3300--10,000 &5-12\\
2007 Aug 21 & 4333.50 & +6.48 & 3300--10,000 &5-12\\
2007 Nov 12 & 4416.50 & +89.48 & 3100--9200  &6\\
2008 Jan 15 & 4480.50 & +153.48&3300--10,000 &5-12\\
  \tableline
  \end{tabular}
  \tablenotetext{a}{Relative to the epoch of $B$-band maximum (JD = 2,454,327.02).}
  \tablenotetext{b}{Approximate spectral resolution (FWHM intensity).}
\end{table}

\end{document}